\documentclass[referee]{aa} 
\usepackage{graphicx}
\usepackage{natbib}
\usepackage{txfonts}
\usepackage{color}

\begin{document}
   \title{Simulation of a flux emergence event and comparison with observations by Hinode}


   \author{L. Yelles Chaouche\inst{1,2}
         \and
         M. C. M. Cheung\inst{3}
         \and
          S. K. Solanki\inst{1,4}
          \and
          M. Sch\"{u}ssler\inst{1}
          \and
          A. Lagg\inst{1}
          }


   \institute{Max-Planck-Institut f\"ur Sonnensystemforschung, Max-Planck-Strasse 2, 37191 Katlenburg-Lindau, Germany
   \and Instituto de Astrof\'{\i}sica de Canarias, C/ V\'{\i}a L\'{a}ctea,
   s/n, E38205 - La Laguna (Tenerife). Espa\~{n}a
   \and Lockheed Martin Solar and Astrophysics Laboratory, Palo
Alto, CA 94304, USA.
   \and School of Space Research, Kyung Hee University, Yongin, Gyeonggi 446-701, Korea\\
              \email{yelles@mps.mpg.de}
             }



  \abstract
  {}
{We study the observational signature of flux emergence in the photosphere
using synthetic data from a 3D MHD simulation of the emergence of a twisted
flux tube.}
{Several stages in the emergence process are considered. At every stage we
compute synthetic Stokes spectra of the two iron lines Fe I 6301.5 {\AA} and
Fe I 6302.5 {\AA} and degrade the data to the spatial and spectral resolution
of Hinode's SOT/SP. Then, following observational practice, we apply
Milne-Eddington-type inversions to the synthetic spectra in order to retrieve
various atmospheric parameters and compare the results with recent Hinode
observations.}
{During the emergence sequence, the spectral lines sample
different parts of the rising flux tube, revealing its twisted structure. The
horizontal component of the magnetic field retrieved from the simulations is
close to the observed values. The flattening of the flux tube in the
photosphere is caused by radiative cooling, which slows down the ascent of
the tube to the upper solar atmosphere. Consistent with the observations, the
rising magnetized plasma produces a blue shift of the spectral lines during a
large part of the emergence sequence.}
  {}

 \authorrunning{Yelles Chaouche et al.}
 \titlerunning{Flux emergence simulation and comparison with observations}
   \maketitle
%

\section{Introduction}

Observational studies of magnetic flux emergence in the photosphere with the
help of Stokes parameters started with the work of \citet{Brants:1985a,
Brants:1985b} and \citet{Zwaan_etal:1985}. Subsequent studies showed that
active regions grow by the successive emergence of magnetic flux fragments,
each containing only a small fraction of the total flux of the active region
\citep{Strous_etal:1996, Strous_Zwaan:1999}. Emerging flux regions (EFRs) of
various sizes lead to magnetic complexes in a range of scales
\citep{Harvey:1993}, from the largest active regions ~\citep[up to $4\times
10^{22}$ Mx;][]{Zwaan:1987} to small ephemeral regions~\citep[with fluxes
down to $10^{19}$ Mx;][]{Hagenaar:1999}.

While the initial studies were restricted to spectropolarimetric observations
of Stokes-$I$, Stokes-$V$ and continuum images \citep{Brants:1985a,
Brants:1985b, Strous:1994}, the full Stokes vector has been used by
\citet[][]{Lites:1998} and \citet[][]{Kubo:2003} for photospheric studies and
by \citet[][]{Solanki:2003nat} and \citet{Lagg:2004, Lagg:2007} for
chromospheric investigations. Recently, observational studies of flux
emergence with high spectral and spatial resolution have been carried out
\citep{Centeno:2007, Cheung:2008, Ishikawa:2008, Martinez:2007,Okamoto:2008}
with the spectropolarimeter \citep[SP;][]{Lites:2001} which is part of the
focal-plane package of the $50$ cm Solar Optical Telescope
\citep[SOT;][]{Tsuneta:2008} onboard the Hinode spacecraft. The emerging flux
regions show horizontally oriented flux elements, which are generally
regarded as the tops of emerging loops. The reported field strength of these
horizontal flux elements varies somewhat among observational studies :
$500\pm 300$ G \citep{Brants:1985a, Brants:1985b}; $200 < |B| < 600$ G
\citep{Lites:1998, Sigwarth:2000}; $400 < |B| < 700$ G \citep{Kubo:2003}; $~
650$ G \citep{Okamoto:2008}. The rising flux elements have an upward velocity
of $\leq 1$ km/s. The estimates of the rise velocity are : $0.5$ km/s
\citep{Brants:1985a, Brants:1985b, Strous_etal:1996}; $\sim 1$ km/s
\citep{Lites:1998}; $< 1$ km/s \citep{Kubo:2003}; 0.3 km/s
\citep{Okamoto:2008}.

\citet{Lites:2005} and \citet{Okamoto:2008} have shown that some flux
elements emerging in the photosphere exhibit a twisted field (cf.
\cite{Wiegelmann:2005}, who found newly emerged loops to harbour considerable
electrical currents).~\citet{Okamoto:2008,Okamoto:2009} (hereafter OK2008 and
OK2009) reported observations of the emergence of a coherent helical flux
tube.

A number of theoretical studies has addressed magnetic flux emergence from
the convection zone to the solar atmosphere. While some approaches treat the
convection zone and overlying atmosphere as plane-parallel hydrostatic layers
\citep[e.g.][etc]{Shibata:1989, Abbett:2003, Archontis:2004, Magara:2006},
others include the effects of convective flows
\citep[e.g.][]{Cheung:2007,Martinez:2008,Isobe:2008}.

In this paper we present for the first time detailed synthetic observational
properties from 3D MHD simulations of an emerging, initially horizontal,
twisted flux tube including its interaction with granular convective flows.
We perform spectral line synthesis and inversion at spectral and spatial
conditions similar to Hinode's SP/SOT. The results are discussed and some
emergence properties are compared with the observations of OK2008 and OK2009,
even though it is clear that there are limits to such a comparison, arising
from the fact that observations and simulations do not occur in the same
solar environment. Nonetheless, we believe that the similarities or
differences revealed by a comparison will provide fresh insight into both
simulations and observations.

\section{Flux emergence simulations} \label{sec:2}

The simulation discussed here has been carried out with the MURaM
code~\citep{Voegler:2005}, which solves the compressible MHD equations taking
into acount non-local energy exchange via non-grey radiative transfer as well
as the effect of partial ionization on the equation of state. Details of the
simulation setup have been described by~\citet{Cheung:2007}. The top and
bottom boundaries are open and allow for mass transfer. The magnetic field is
matched to a potential field at the top boundary. The initial condition
consists of a horizontal buoyant magnetic flux rope (i.e., a flux tube with a
twisted magnetic field) embedded in the convection zone at a depth of $1.35$
Mm below the solar surface. The initial axial field strength and
characteristic radius of the tube are $B_0 = 8500$ G and $R_0=200$ km,
respectively, which gives a longitudinal flux of $\Phi = 10^{19}$ Mx. The
dimensionless twist parameter of the tube (corresponding to the ratio of the
transverse and longitudinal field strengths at a distance $r=R_0$ from the
tube axis) is $\lambda=0.5$. This setup corresponds to run U1
of~\citet{Cheung:2007}. To ensure that the temperature structure in the
photosphere is sufficiently realistic for the synthesis of Stokes parameters,
we have repeated this run with non-grey radiative transfer using four opacity
bins \citep{Voegler:2004}.

As shown by \citet{Cheung:2007}, the setup described above allows a magnetic
tube with a flux of $10^{19}$ Mx to overcome the downward directed drag of
convective downflows and emerge as a coherent structure, although the effects
of the convective pattern are still impressed upon the emerging flux pattern
at the solar surface.

\section{Spectral line synthesis and Milne-Eddington inversion} \label{sec:3}

\begin{figure*} 
\includegraphics[width=0.87\textwidth,bb= 65 125 450 380]{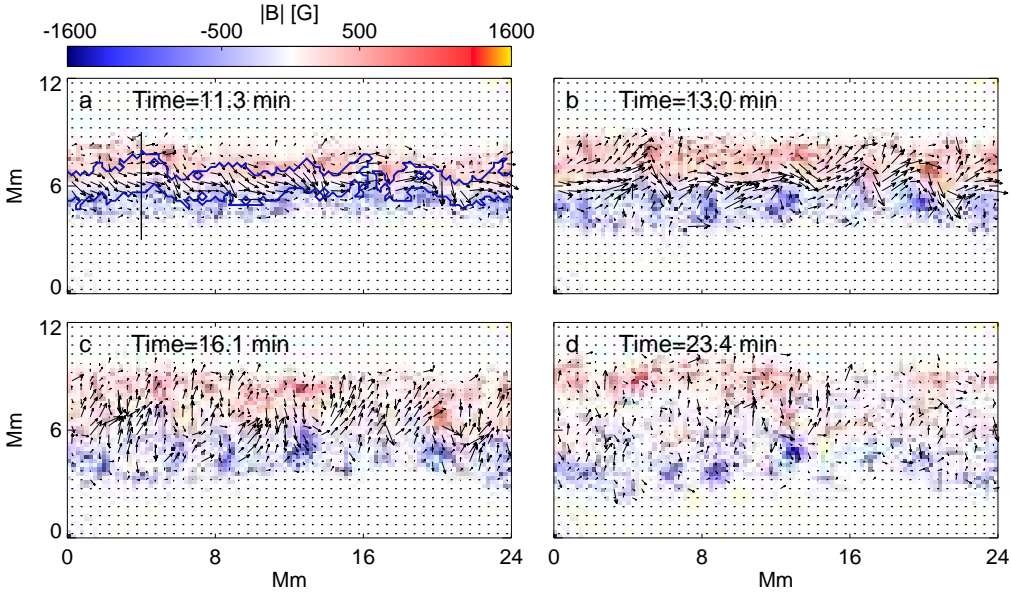}
\caption{Maps of the magnetic field vector retrieved from Stokes inversions.
The colour code indicates the amplitude of the vertical (line-of-sight)
component of the magnetic field. The orientation of the horizontal field
components is shown by arrows. The panels a to d correspond respectively to:
$t=11.3, 13.0, 16.1$ and $23.4$ min. The blue contour in the upper-left panel
outlines regions where the horizontal component of the magnetic field is
larger than 300 G. The vertical black line in the upper-left panel indicates
the location where vertical cuts in the MHD cubes are plotted in
Figs.~\ref{fig2} and~\ref{fig3}} \label{fig1}
\end{figure*}

We have computed the Stokes parameters for the iron lines Fe I $6301.5$ {\AA}
and Fe I $6302.5$ {\AA}. This has been done in LTE using the STOPRO code in
the SPINOR package \citep{Solanki:1987, Frutiger:2000}. For the atomic
parameters characterising these two lines and the iron abundance we have used
the same values as \cite{Shchukina:2001};
\cite{Khomenko:2005a,Khomenko:2005b}; \cite{Borrero:2006} and
\cite{Shelyag:2007}.

The resulting Stokes data are degraded to Hinode's SP/SOT resolution
employing an optical model of the telescope and taking into account a defocus
value typical of the Hinode's SP following \citet{Danilovic:2008}. We use a
pixel size of 0.32 arcsec. In line with typical characteristics of Hinode
SP/SOT, we degrade the Stokes spectra to a spectral resolution of 30 m{\AA}
and a spectral sampling of 21.6 m{\AA}. A photon noise of $10^{-3}$ $I_{c}$
is also added.

The degraded synthetic Stokes data are inverted using a Milne-Eddington
inversion procedure. This is done by using the Helix code \citep{Lagg:2004}.
Both spectral lines are inverted simultaneously. The existence of two
atmospheric components within each pixel was allowed for, one harbouring a
magnetic field, the other being field-free. The fraction of the surface area
occupied by a magnetic field within each pixel is the "filling factor". The
inversion procedure searches for the optimal combination of parameters
corresponding to the minimum of the merit function. The following free
parameters are considered: the magnetic field strength, its inclination and
azimuth, the ratio of line center to continuum opacity, the Doppler width,
the damping constant, the slope and surface value of the source function, the
Doppler shift (giving the line-of-sight velocity) and the filling factor.

\section{Results and discussion} \label{sec:4}

Figure~\ref{fig1} shows a time sequence of the magnetic field maps retrieved
by the inversion. Clearly, the emerging field expands laterally with time.
Similar expansion has been reported in observations of the emergence of a
flux rope in an active region by OK2008 and OK2009 (see Fig. 2 in OK2008
during the first half of the rise of the observed flux rope).

The orientation of the horizontal components of the field is indicated by
arrows. The twist of the emerging flux tube is revealed by the orientations
of the field lines, although the apparent structure is deformed by the
influence of the granulation. This orientation changes along the emergence
sequence (see the difference between the two snapshots at $t=11.3$ min and
$t=16.1$ min).

At $t=11.3$ min the spectral lines sample the upper part of the emerging flux
tube, where the field lines are predominantly oriented from upper left to
lower right in the plotted frame. Later, at $t=16.1$ min, the flux tube has
emerged further and the spectral lines sample its lower part, where the field
lines in this twisted structure are oriented from the lower left to the upper
right. A change in the field orientation has been reported in the observed
event by OK2008 and OK2009 although the observed event takes place in an
active region.

The mean value of the horizontal component of the magnetic field in the
region outlined with the blue contour (near the polarity inversion line) in
the upper left panel of Fig.~\ref{fig1} is $550$ G at $t=11.3$ min. This
value is close to the one ($650$ G) reported by OK2008 and OK2009. The
horizontal field strength exhibits higher values in the close vicinity of the
polarity inversion line and decreases with time along the emergence sequence
shown in Fig.~\ref{fig1}. The upward velocity of the emerging flux rope in
the same region outlined in blue is $670$ m/s. This value is larger than the
$300$ m/s reported in OK2008 and OK2009 and indicates a faster emergence rate
of the simulated flux rope. The mean value of the magnetic field inclination
with respect to the vertical in the region outlined in blue is nearly zero
degrees (zero corresponds to a horizontal field) with a standard deviation of
about 30 degrees. In OK2009, it is reported that the mean inclination near
the polarity inversion region is nearly zero, with a scattering of the order
of $\pm 10$ degrees. The mean vertical magnetic field in the same outlined
region is nearly zero. The one reported in OK2009 is 0 $\pm 200$ G.


The total unsigned magnetic flux through the maps in Fig.~\ref{fig1} is
calculated from the inversion results as: Flux = $A \displaystyle\sum_{i}
|B_{z,i}| f_{i}$, where $A=$ pixel area, $B_{z,i}=B_{z}$ in pixel $i$ and
$f_{i}=$ filling factor in pixel $i$. For the four snapshots a-d in
Fig.~\ref{fig1} we obtain the following values: $1.15 \times 10^{20}$ Mx,
$1.42 \times 10^{20}$ Mx, $1.45 \times 10^{20}$ Mx and $1.05 \times 10^{20}$
Mx, respectively. The increase of the unsigned flux in the first three
snapshots is due to flux emergence. Subsequently, the decrease is due to flux
cancellation of opposite-polarity flux elements. These elements are formed by
the fragmentation of the tube under the action of flux expulsion
\citep{Voegler:2005}. The encounter of opposite polarity flux elements in the
intergranular lanes leads to the decrease of the unsigned flux.

\begin{figure} 
\includegraphics[width=0.9\textwidth,bb= 75 123 760 590]{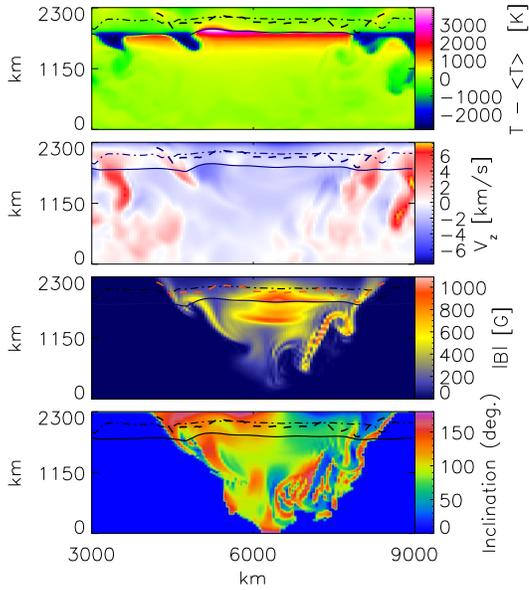}
\caption{Vertical cut through the simulation box at the location shown by the black
line in the upper left panel in Figure~\ref{fig1} at $t= 11.3$ min. The
panels represent, from top to bottom, the temperature fluctuations with
respect to the horizontal average, vertical velocity (blue represents
upflow), magnetic field strength and magnetic field inclination. The solid black lines indicate the
location of $\tau_{5000}=1$. The dash-dotted lines show the height at which
the contribution function of Stokes-$I$ reaches its maximum, and the dashed
lines indicate the same for Stokes-$V$.} \label{fig2}
\end{figure}

2D vertical cuts through the MHD cubes perpendicular to the axis of the
original flux rope are shown in Figs.~\ref{fig2} and~\ref{fig3} at t=11.3 min
and 16.1 min, respectively. They lie along the horizontal coordinates
represented by the black line in the upper left panel of Fig.~\ref{fig1}.

The rising flux tube flattens considerably after reaching the solar surface.
The upper parts of the flux tube (above the solar surface) rapidly cool down
radiatively. Therefore, the plasma becomes denser and loses its buoyancy. On
the other hand the bulk of the flux tube keeps rising. This is due to the
fact that the rising sub-surface parts of the flux tube keep pushing the
plasma upward above the surface \citep{Cheung:2007}. The material above the
surface also expands laterally due to the strong stratification of the
photospheric layers and partly descends back into the convection zone at the
sides of the flux tube. We see a blue shift during most of the emergence
sequence at the height where Fe I 6301.5 {\AA} and Fe I 6302.5 {\AA} are
formed, except for downflow lanes along the flux rope where we find a red
shift \citep[][]{Cheung:2007,Yelles:2008}. A blue shift has been observed as
well by OK2008 during a large part of the emergence sequence.

Figure~\ref{fig2} shows that the flux rope has a relatively regularly twisted
core with a field strength approaching 1 kG near and below the solar surface.
Note that the Stokes-$V$ signal is formed over a height-range sampling mainly
sub-kG field. Convective flows affect the outer parts of the emerging flux
tube more strongly, so that the field configuration becomes more complex than
in the central part. This is best seen in the field inclination (lowest panel
of Fig.~\ref{fig2}) and is mainly restricted to the sub-surface layers. At
the height of the line formation the field is rather homogeneous and the
opposite polarities are separated (see also the top left panel in
Fig.~\ref{fig1}).


\onlfig{3}{
\begin{figure} 
\includegraphics[width=0.9\textwidth,bb= 75 123 760 590]{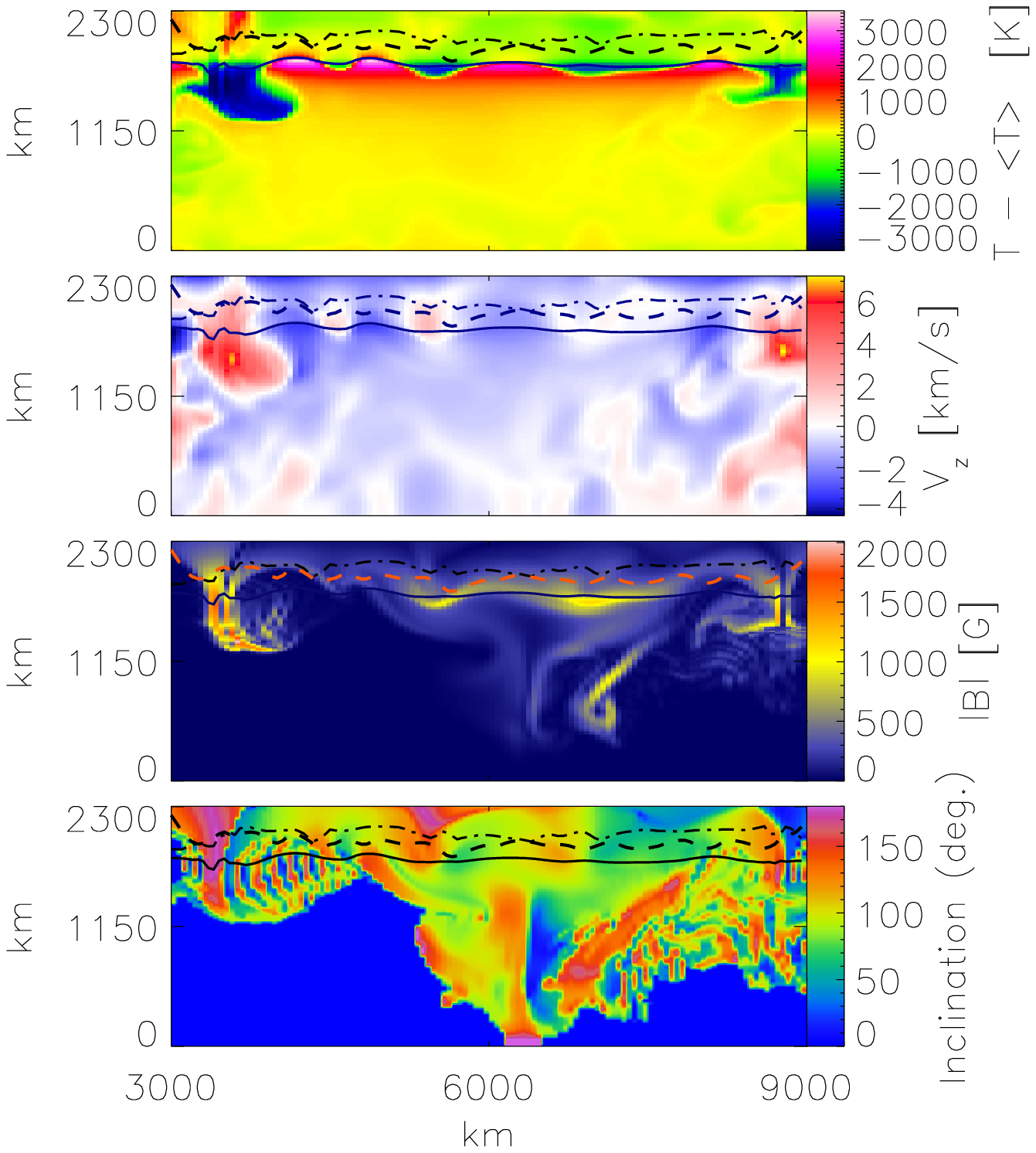}
\caption{Similar to Figure~\ref{fig2} at $t= 16.1 $ min.} \label{fig3}
\end{figure}
}

At $t=16.1$ min (Fig.~\ref{fig3}), one can still see the downflows beside the
flux tube, where the field is predominantly vertical. The flux tube has
become wider, and there are patches of downflowing plasma above the solar
surface in the body of the flux tube (e.g. near the abscissa $x=5500$ km).
The horizontally oriented core of the flux tube has greatly shrunk. Part of
the flux has been advected to downflow lanes and concentrated to kilo-Gauss
field strength. This produces a strong circularly polarized signal correlated
with downflow lanes in the emergence area. At this point in time the flux
rope is starting to loose its coherence even at the height of the line
formation (Fig.~\ref{fig3}), with multiple islands of both polarities being
present.

By the time of the last plotted snapshot (Fig.~\ref{fig1}d) many islands of
opposite polarities are visible. Nonetheless, a predominance of positive
polarities in the upper and negative polarities in the lower part of the
frame can be discerned. In vertical cuts similar to Figs.~\ref{fig2}
and~\ref{fig3} we show how the inhomogeneity of the magnetic structure slowly
propagates upwards.

\section{Discussion} \label{sec:5}

We have studied the observational signature of a twisted flux tube emerging
through the solar photosphere with the help of 3D MHD simulations. This
emergence event reproduces some of the characteristics of the event studied
by OK2008 and OK2009.

There are also major differences, however, which we now briefly discuss:
Whereas the channel of emerging flux closes on itself according to OK2008 we
find no sign of this in our simulations. Instead, the emerged field at the
surface continues to expand. At the same time the structure of the vertical
component of the field becomes increasingly inhomogeneous. This is similar to
the evolution of an emerging loop system studied by \citet{Cheung:2008}.
Also, the simulated flux rope emerges within a few tens of minutes whereas
the observed emergence takes place over roughly a day.


\cite{Cheung:2007, Cheung:2008} have made a parametric study where they
discuss the dependance of the emergence properties on the initial magnetic
field strength and twist. The speed of emergence depends on the total
buoyancy of the tube (related to total flux and total density deficit). In
that sense, more buoyant tubes rise more quickly and can modify the downflow
pattern. Very buoyant and twisted tubes tend to modify the granulation
pattern in an obvious way that is lacking in the OK2008 event. We can then
assume that the buoyancy of the observed emerging flux is not very large,
causing no anomalous granulation and has gentle flows. This is consistent
with the fact that the observed emergence has an upflow speed lower than the
simulated case and takes place over a longer period. We know the behaviour of
the simulated emergence only over a rather limited range of parameter space
(initial magnetic field strength and twist) studied by \cite{Cheung:2007,
Cheung:2008}, which is still different from the regime of the OK2008 case. In
future work, we plan to investigate flux emergence leading to and within
active regions. The emergence process inside a pre-existing field will be of
particular interest.

\begin{acknowledgements}

We are grateful to S. Danilovic and A. Gandorfer for providing the Hinode
optical model. This work was partly supported by the WCU grant No. R31-10016
from the Korean Ministry of Education, Science and Technology. L.Y.C. is
thankful to the Max-Planck-Institut f\"{u}r Sonnensystemforschung,
Katlenburg-Lindau, for a post-doctoral stipend.

\end{acknowledgements}

\bibliographystyle{aa}
\bibliography{main}
\end{document}